\documentclass[twocolumn,showpacs,preprintnumbers,superscriptaddress,prl,floatfix,aps,10pt]{revtex4-1}

\usepackage{subfigure}
\usepackage{graphicx}
\usepackage{dcolumn}
\usepackage{bm}
\usepackage{color}
\usepackage{amssymb}
\usepackage{amsmath}
\usepackage{epsf}
\usepackage{epstopdf}%
\usepackage{amsfonts}
\usepackage{float}
\restylefloat{table}

\usepackage[english]{babel}

\usepackage{todonotes}
\definecolor{antoine}{RGB}{192,225,215}
\definecolor{zz}{RGB}{90,150,175}
\definecolor{matt}{RGB}{250,160,130}

\begin{document}


\title{Spectroscopic properties of few-layer tin chalcogenides}

\author{Antoine Dewandre}
\affiliation{Department of Physics, Universit\'e de Li\`ege, all\'ee du 6 ao\^ut, 19, B-4000 Li\`ege, Belgium.}
\affiliation{European Theoretical Spectroscopy Facility www.etsf.eu}

\author{Matthieu J. Verstraete}
\affiliation{Department of Physics, Universit\'e de Li\`ege, all\'ee du 6 ao\^ut, 19, B-4000 Li\`ege, Belgium.}
\affiliation{European Theoretical Spectroscopy Facility www.etsf.eu}

\author{Nicole Grobert}
\affiliation{Department of Materials, University of Oxford, Parks Road, Oxford OX1 3PH, UK}
\affiliation{Williams Advanced Engineering, Grove, Oxfordshire, OX12 0DQ, UK}

\author{Zeila Zanolli}
\affiliation{European Theoretical Spectroscopy Facility www.etsf.eu}
\affiliation{Catalan Institute of Nanoscience and Nanotechnology, CSIC and BIST, Campus UAB, Bellaterra, 08193 Barcelona, Spain}

\begin{abstract}
Stable structures of layered SnS and SnSe and their associated electronic and vibrational spectra are predicted using first-principles DFT calculations. The calculations show that both materials undergo a phase transformation upon thinning whereby the in-plane lattice parameters converge to a pseudo-cubic phase, similar to the high-temperature behaviour observed for their bulk counterparts. The electronic properties of layered SnS and SnSe evolve to an almost symmetric dispersion whilst the gap changes from indirect to direct. Characteristic signatures in the phonon dispersion curves and surface phonon states where only atoms belonging to surface layers vibrate can also be observed for these materials.

\end{abstract}

\maketitle


\section{Introduction}
Chalcogenides are a remarkable family of layered materials, displaying an extensive range of optical, electronic, thermal and mechanical effects \cite{Eggleton_2011}. They are used as phase-change materials in rewritable data storage \cite{Wuttig_yamada_2007}, as high performance thermoelectrics \cite{Zhao_2014,dewandre_2016}, and as absorbing layers in photovoltaic cells \cite{Bur-13, Mal-13, Vid-12}. 
Lead (Pb) chalcogenides and their alloys have been heavily studied for their excellent thermoelectric properties\cite{lalonde_2011, lalonde_reevaluation_2011, romero_2015, shafique_thermoelectric_2017} but the presence of toxic chemical elements is a major industrial disadvantage. Great efforts have been invested recently in a less toxic analogue: Tin chalcogenides (SnX with X=S, Se, Te).

Researchers are currently investigating emerging properties in 2D confined systems aiming at their exploitation in a wide variety of applications  \cite{Nov-04, Kos-13, Col-11, Wan-12}.
Monochalcogenides are naturally layered compounds, which can be grown in the form of flakes only a few mono-atomic layers thick\cite{li_revealing_2017}. The reduction of the dimensionality of the crystal (3D to 2D material) has an impact on the geometry of the layers and hence on elastic, electronic and vibrational properties, which vary with the number of  layers. Recent theoretical studies of electronic structure changes in few-layer SnX have been reported, indicating that the band gap expands significantly (from 1.32 eV indirect to 2.72 eV direct) as the number of layers is decreased\cite{Tri-13}. Deb and Kumar reproduced this study taking into account the relaxation of the atomic positions and the unit cell \cite{Deb_2017}. Mehboudi \textit{et al.} also studied electronic and optical properties for mono and bilayer\cite{Mehboudi_2016}.

In what follows, the structural, vibrational and electronic properties of SnS and SnSe are studied to gain insight into the behaviour of these compounds in few-layer form. In particular, we focus on the possible structural transformations, variations of the vibrational spectra and electronic structure resulting from reduction of the dimensionality of the crystal. Both Raman spectroscopy and reflectivity measurements offer easy and non-destructive characterization methods: a comparison with the computationally predicted spectra can be used to assess the material thickness. We represent the results as a function of \textit{1/n} where \textit{n} is number of layers. This gives us a easy way of including the bulk properties ($n = \infty$) and visually appreciate the convergence of the different properties towards this limit.

\section{Few-layer structures}

\begin{figure}[htbp]
\centering
\includegraphics[width=0.29\textwidth]{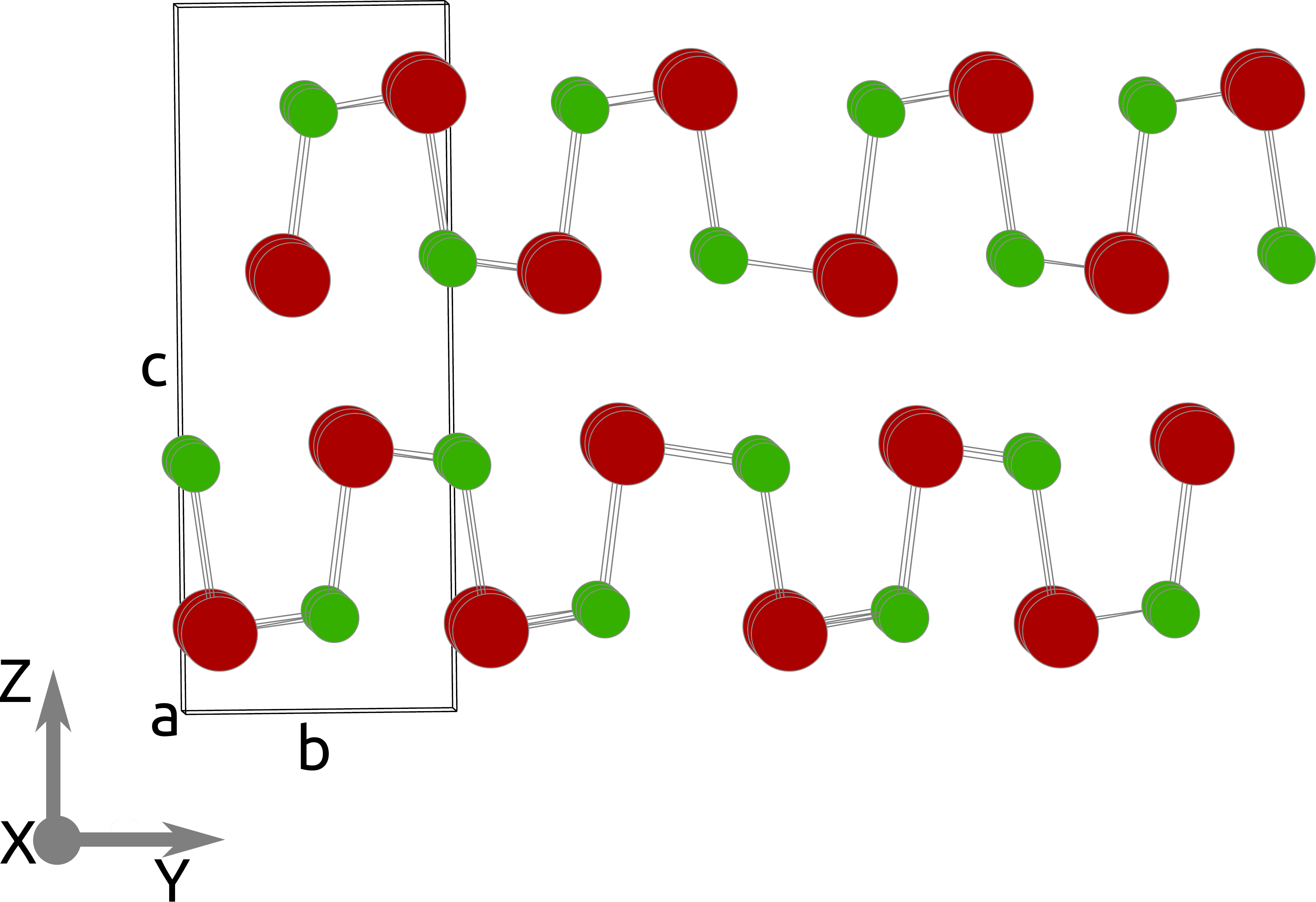}
\includegraphics[width=0.2\textwidth]{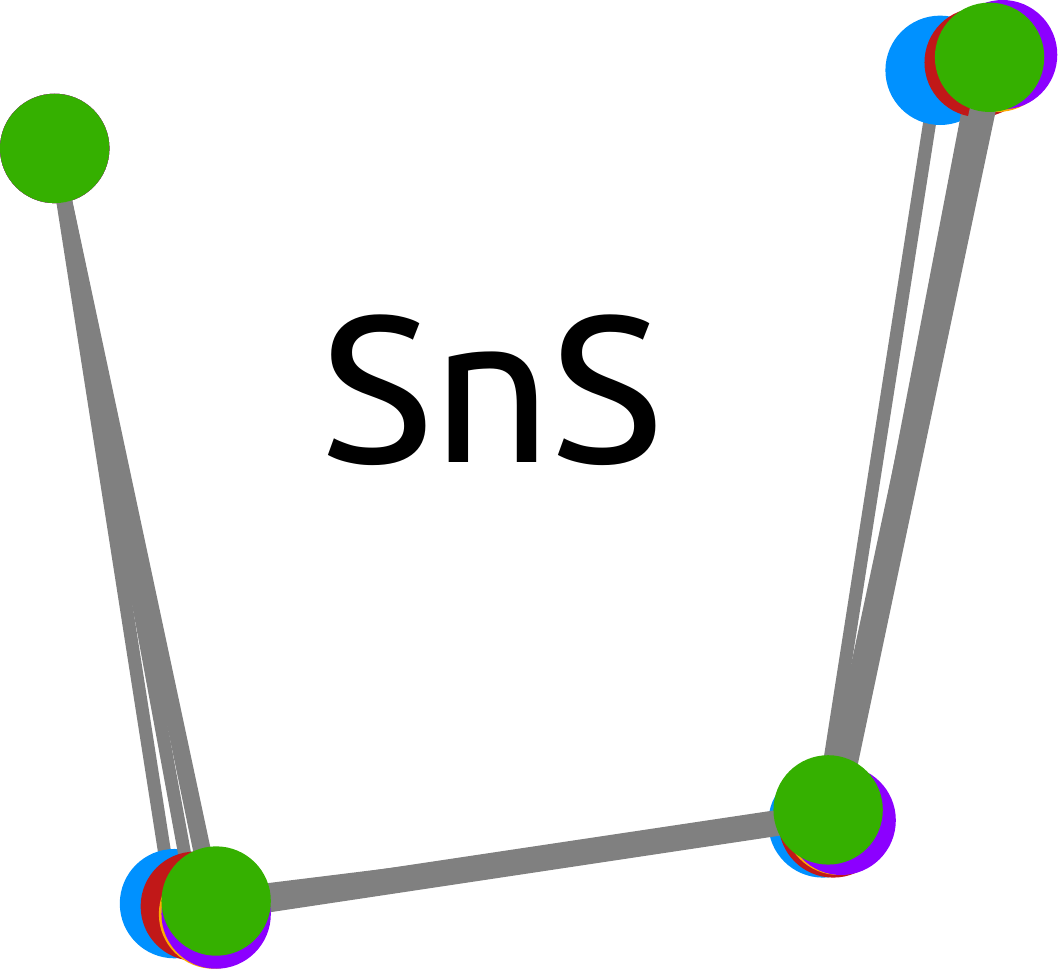}
\includegraphics[width=0.2\textwidth]{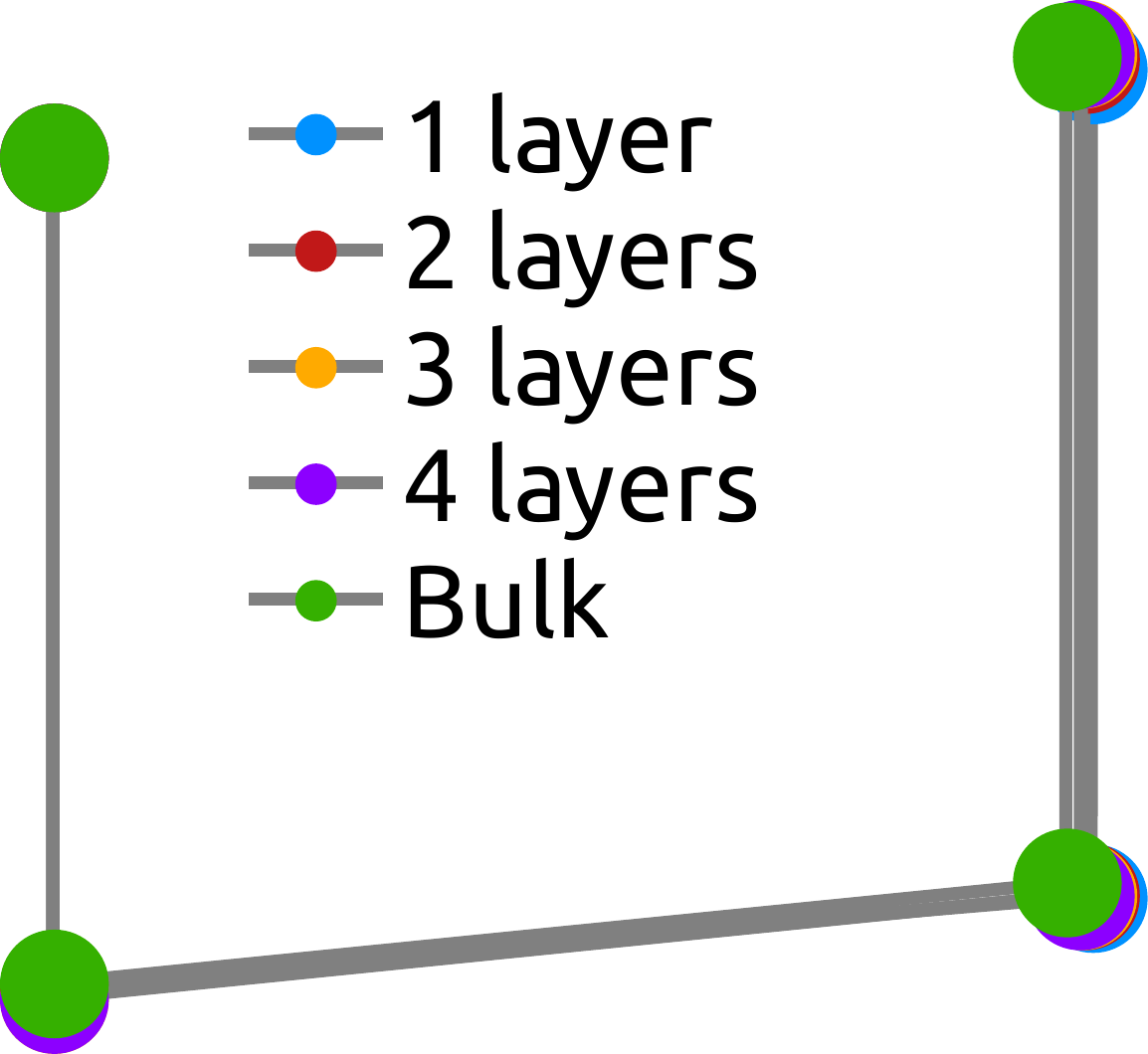}
\includegraphics[width=0.2\textwidth]{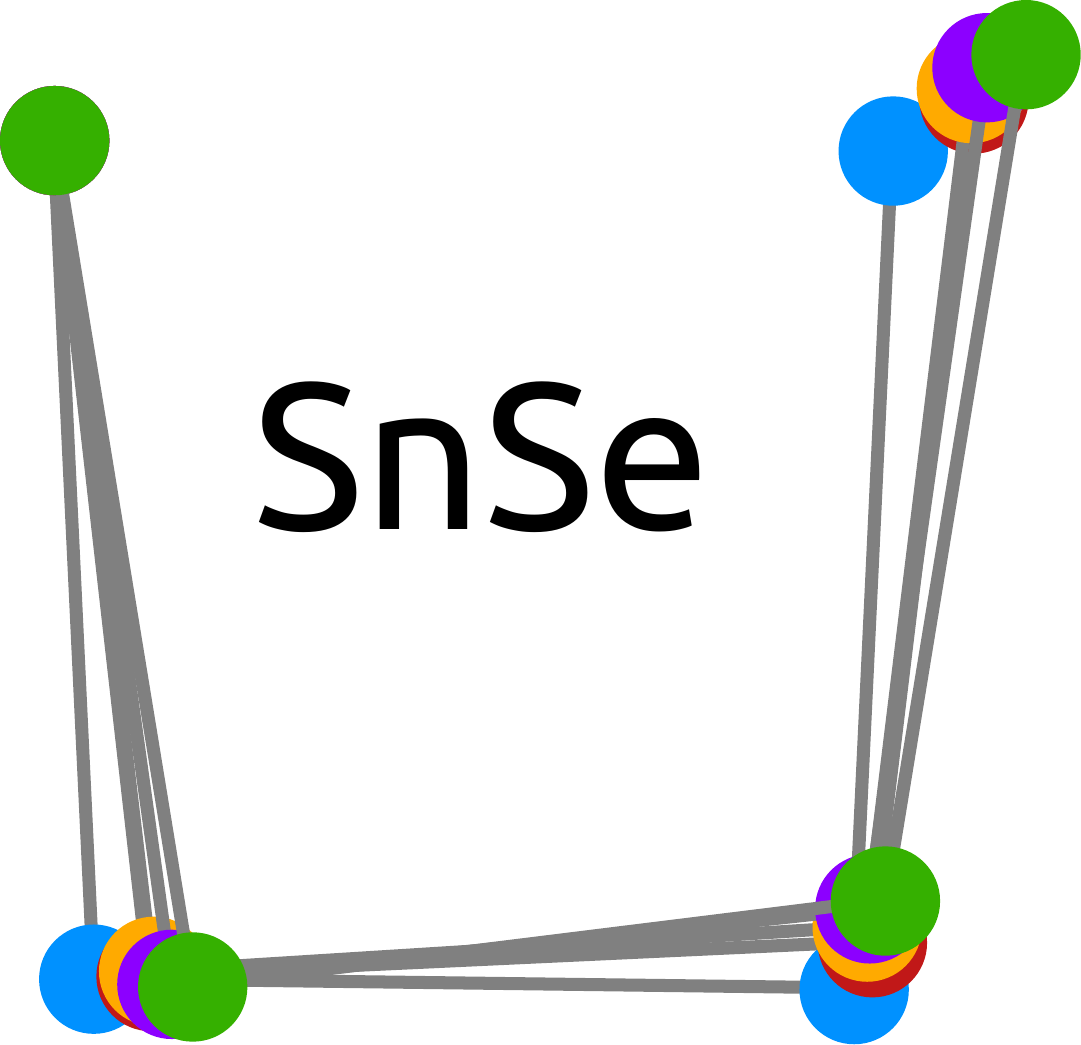}
\includegraphics[width=0.2\textwidth]{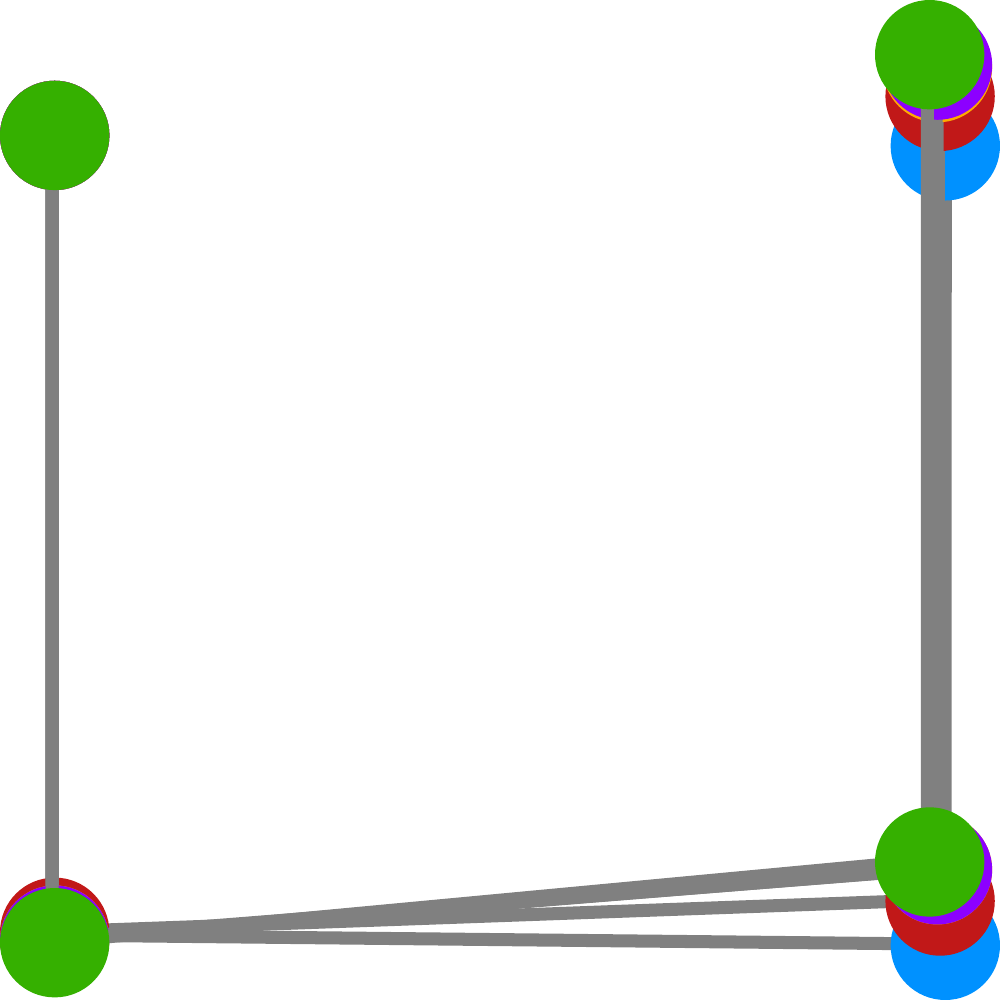}
\caption{Layered orthorhombic ({\it Pnma}) bulk crystal structure of  SnX (X = S, Se): red and green circles represent Sn and X, respectively. Projection of the YZ (left) and XZ (right) planes for both compounds (slabs and bulk) shifted to align the atoms in the upper left corner. In the YZ plane, the accordion-like bonds are stretched as the number of layers are increased. The angle departs from 90$^\circ$ and the layer evolves towards the {\it Pnma} positions as the number of layers increases. The puckering of the surface differs when comparing SnS and SnSe. In SnSe, the height of the upper atoms changes, creating a puckered surface, while in SnS the difference in relative height of the atoms remains constant with the number of layers.}
\label{sns_structures}
\end{figure}

Tin chalcogenides exist in an orthorhombic ({\it Pnma}) bulk structure comprising weakly coupled layers of covalently bound Sn-X (X = S or Se) atomic bilayer units (Figure \ref{sns_structures}). Both materials can be isolated in few-layer form \cite{jiang_2017}. In the following, "layer" will refer to the natural atomic bilayer unit. 
We consider the structural distortions appearing when this compound is isolated in free standing slabs, with thicknesses from 1 to 6 layers, and compare the geometry of the slabs with the bulk.
The binding energy of SnS and SnSe slabs are 30 and 10 meV/\AA$^2$ respectively (see supplemental information) which classifies these compounds as ``easily/potentially exfoliable'' according to the criteria used in Mounet et al.~\cite{marzari_2016}.

We perform Density Functional Theory (DFT) calculations where the exchange correlation terms are calculated within the Generalized Gradient Approximation (GGA) of Perdew et al.\cite{perdew96}. The structures of Sn-X bulk are obtained by fully relaxing the internal positions and the shape of the unit-cell. Relaxed and experimental lattice constants are specified in table \ref{table_latticeconstants}. These results slightly overestimate (within 2\%) the experimental values, which is within the error expected for semi-local DFT calculations with the GGA.

\begin{table}[H]
\begin{center}
\begin{tabular}{cc|ccc}
• & • &  \textit{a} [ \AA ]  &  \textit{b} [ \AA ]  &  \textit{c} [ \AA ]  \\ 
\hline 
SnS   &  Computed  &  4.06  &  4.41  &  11.40  \\ 

• & Exp  \cite{vonschnering_1981} &  3.98  &  4.33  &  11.18  \\ 
\hline 
SnSe  &  Computed  &  4.21  &  4.50  &  11.72  \\ 

• &  Exp  \cite{Zhao_2014}  &  4.13  &  4.44  &  11.49  \\ 
\end{tabular} 
\end{center}

\caption{Relaxed and experimental lattice parameters of bulk Sn-X. Computed results overestimate the experimental ones by 1 to 2 \%, which is expected in DFT-GGA.}
\label{table_latticeconstants}
\end{table}

\begin{figure}[htbp]
\centering
\includegraphics[width=0.45\textwidth]{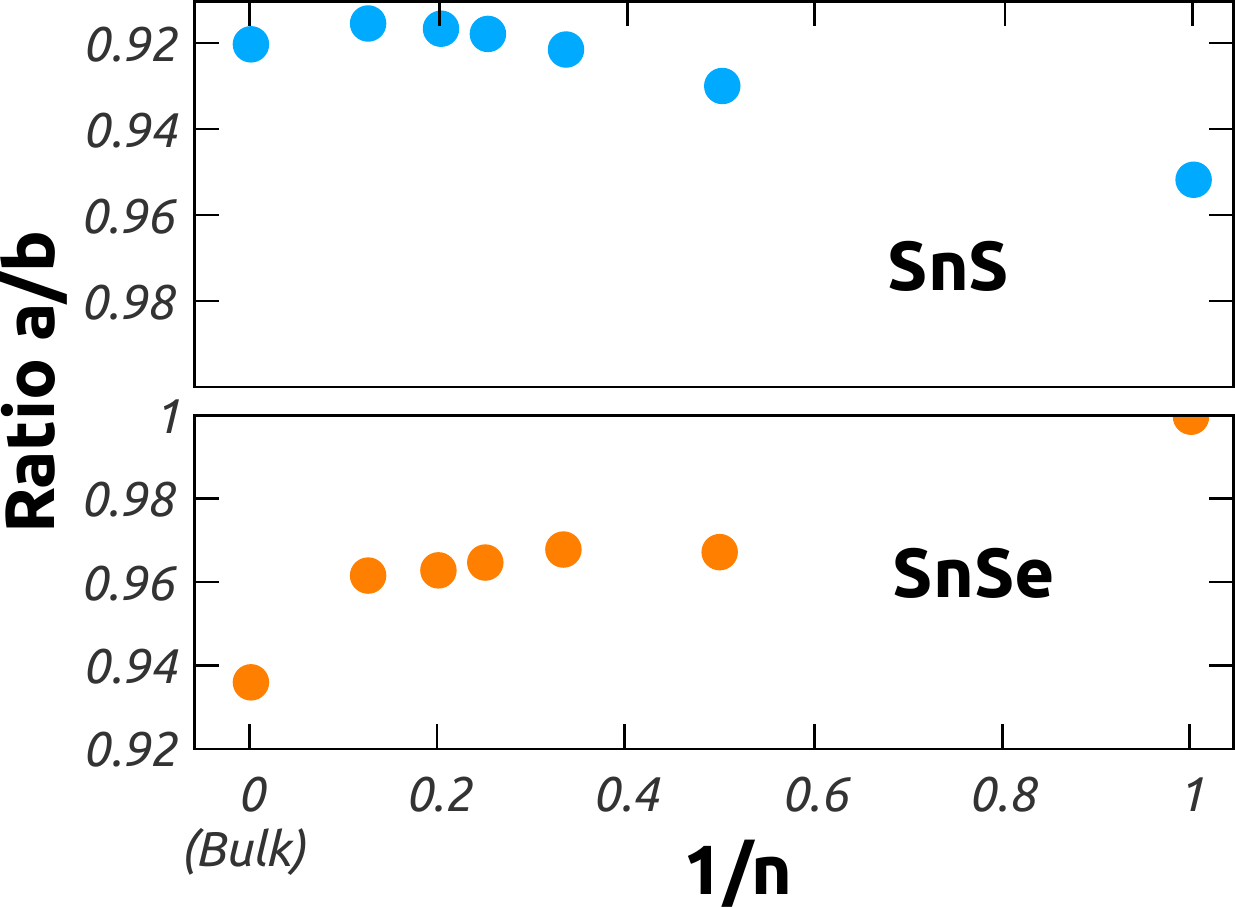}
\caption{\textit{a/b} ratio as a function of the inverse of the number of layers.}
\label{ratio}
\end{figure}

Few-layer structures are relaxed, and the lattice parameters are allowed to vary in the in-plane directions. Details of the calculations can be found in the Methods section. As can be seen in Fig.~\ref{ratio}, when reducing the number of layers, the in-plane lattice parameters (\textit{a} and \textit{b}) converge, and are almost identical for the monolayer case of SnSe. This behaviour mirrors experimental studies on heterostructures containing SnSe slabs \cite{Ludemann_2014}. The monolayer of SnS also rectifies, but its \textit{a/b} ratio does not converge to 1: we find a small dynamical instability for square in-plane parameters (see Supplemental information). Condensing these unstable modes at $\Gamma$ creates a distortion on the atomic coordinates along the \textit{a} axis, breaks the symmetric 0.25/0.75 reduced coordinates, and leads to $a/b \neq 1$. The resulting symmetry of the crystal is triclinic. The difference in total energy between the two structures is 9 meV/f.u., and it is entirely possible that epitaxy or other substrate constraints will stabilize the square lattice of the SnS monolayer as well.

The internal coordinates and the interlayer distance of the atoms also evolve with the slab thickness. To visualize the evolution of the internal coordinates of the atoms, their projections on the YZ  and XZ planes have been superimposed in Fig. \ref{sns_structures}. The positions of the upper left atoms have been aligned and only the innermost layer of the slab is shown.
The YZ projection shows that the bond angles deviate from 90$^\circ$ when the number of layers increases, to acquire the familiar accordion shape of bulk. The change in the XZ plane is much weaker. 
The internal coordinates show surface effects already in the 3 layer case, and inner layers converge towards the shape of the bulk (Fig. \ref{sns_structures}) as the thickness of the slab increases. Surface layers converge, instead, towards coordinates between those of the bulk and the monolayer.

\begin{figure}[htbp]
\centering
\includegraphics[width=0.45\textwidth]{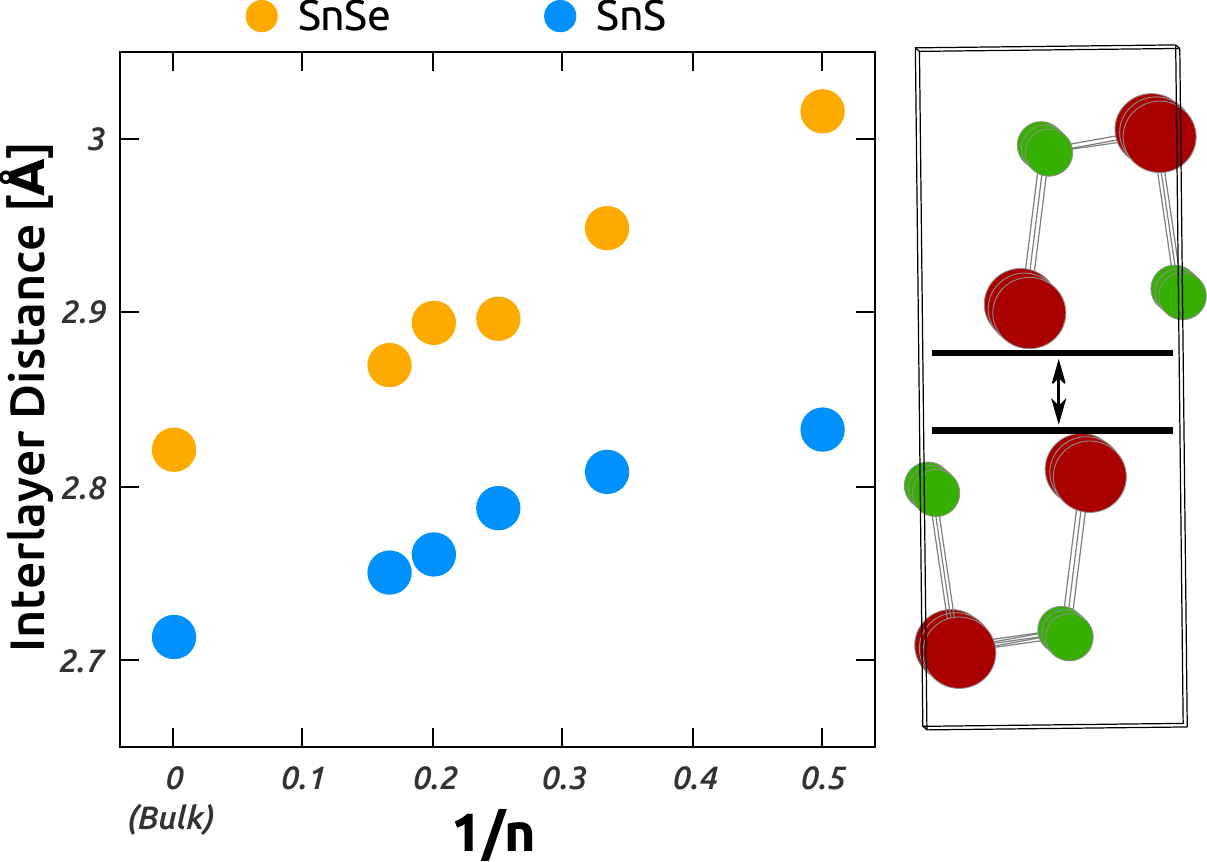}
\caption{Interlayer distance between of SnX layers. Right panel: the distance is measured between the atoms closest to each other in adjacent layers (Sn atoms).}
\label{interlayer}
\end{figure}

Because of the attractive electrostatic interaction between the layers, the interlayer distance decreases with the number of layers following a \textit{1/n} law (Fig. \ref{interlayer}). The interlayer gap is significantly smaller in the SnS case, and varies less.  
The interlayer distances are measured between closest atoms belonging to adjacent layers as depicted in the right panel of Fig. \ref{interlayer}. Starting from 4 layers, differences can be seen between those on the edge and those in the center of the slab. The distances reported in Fig.~\ref{interlayer} correspond to the central ones.

We notice a subtle difference between the two compounds: the puckering of the surface (Sn and X atoms are not on the same plane of the surface of a layer) slowly disappears when reducing the number of layers in SnSe while it stays constant in SnS. This difference shows distinct hybridization of the (on average) half-filled p orbitals at the surface, which is closer to pure $p_{x,y,z}$ in SnSe, leading to a quasi-cubic structure, as in $\alpha$-Po \cite{legut_2007_Po_SC_relativity, verstraete_2010_polonium} or high pressure Ca \cite{digennaro_2013_calcium}. This may be linked to the size of the surface lone pairs, or to the relative alignment of the orbital energy levels.


\section{Electronic bandstructures}

Electronic band structures were calculated for bulk and few-layer materials using the relaxed lattice parameters reported above. Comparison with experimental bulk values shows an underestimation of the fundamental gap, which is expected for DFT calculations, but can often be corrected with a constant factor. The fundamental gap was measured by Albers \textit{et al.} at 1.08 eV \cite{albers_1961} for SnS and 0.90 eV \cite{albers_1962} for SnSe, while our results yield 0.85 eV and 0.57 eV respectively. 

\begin{figure}[htbp]
\centering
\includegraphics[width=0.45\textwidth]{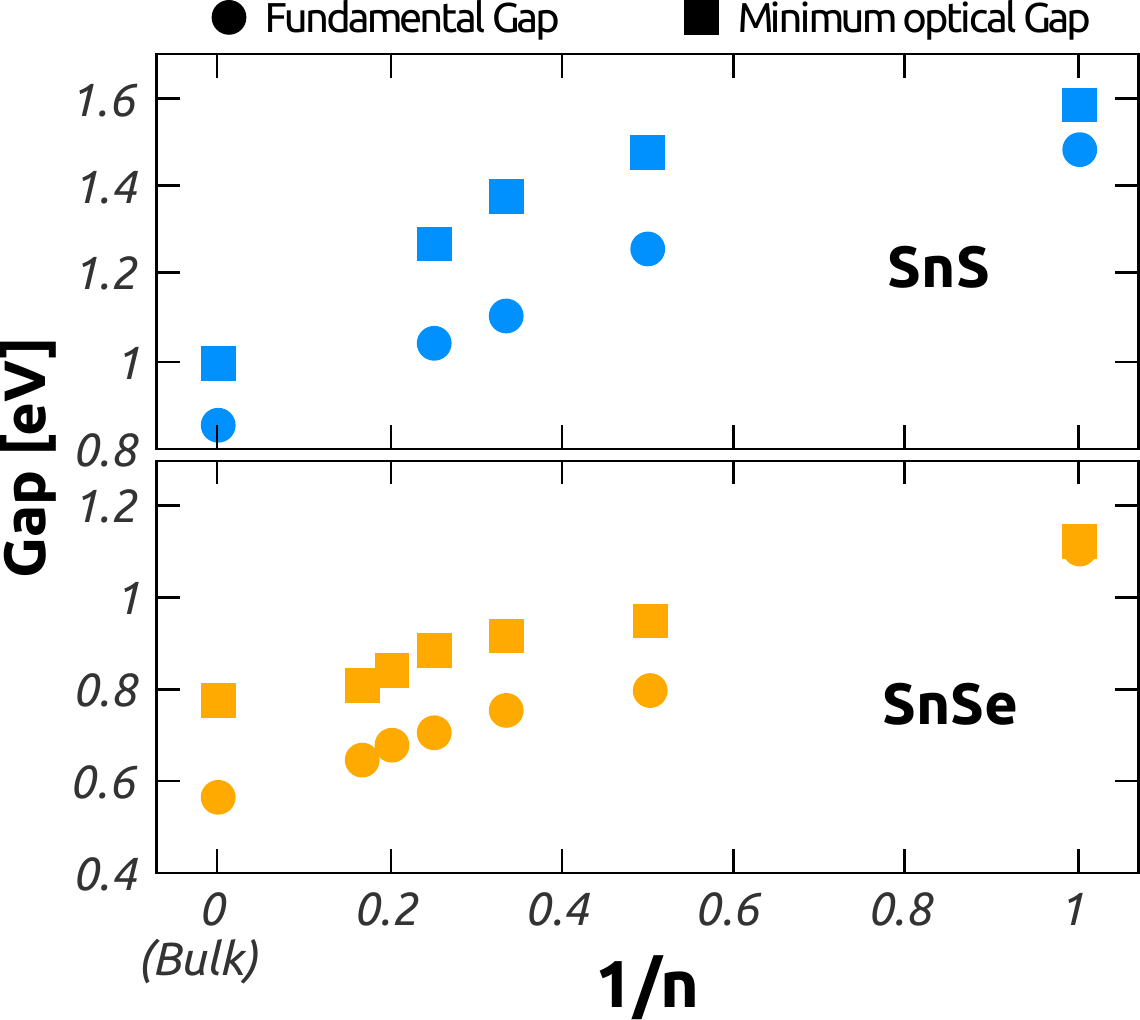}
\caption{Electronic gaps of the multilayer and bulk SnX compounds, as a function of the inverse of the number of layers $n$. The gap is proportional to $1/n$. In both compounds, the gap of the monolayer is almost direct.}
\label{gap}
\end{figure}

Fig.~\ref{gap} shows the fundamental and optical gaps of the SnX compounds as a function of the number of layers. The band gap increases for thinner layers, due to quantum confinement effects. Nevertheless, the fundamental gap remains indirect in our calculations, except for the monolayer were the gap is almost direct (the difference between fundamental and optical gap is less than 0.03 eV). The reduced energy difference between indirect and direct band gaps will result in a sharper onset of optical absorption.

A comparison between our band structures for relaxed lattice parameters and fixed bulk lattice parameters \cite{Tri-13}, shows that the X-M-Y path presents a higher degree of symmetry in the relaxed case. This is directly related to our prediction that the in-plane lattice parameters converge as the number of layers is reduced, and the associated atomic rearrangement tends towards a more cubic structure. Globally, we are able to tune the band gap over a large range ( 0.6 and 0.4 eV for SnS and SnSe, respectively)  by changing the thickness of these compounds, which is very useful in optoelectronic applications. The electronic dispersions for both compounds can be found in supplemental information.

\section{Raman and Reflectivity Spectra}

\begin{figure}[htbp]
\hspace{2em}
\includegraphics[width=0.45\textwidth]{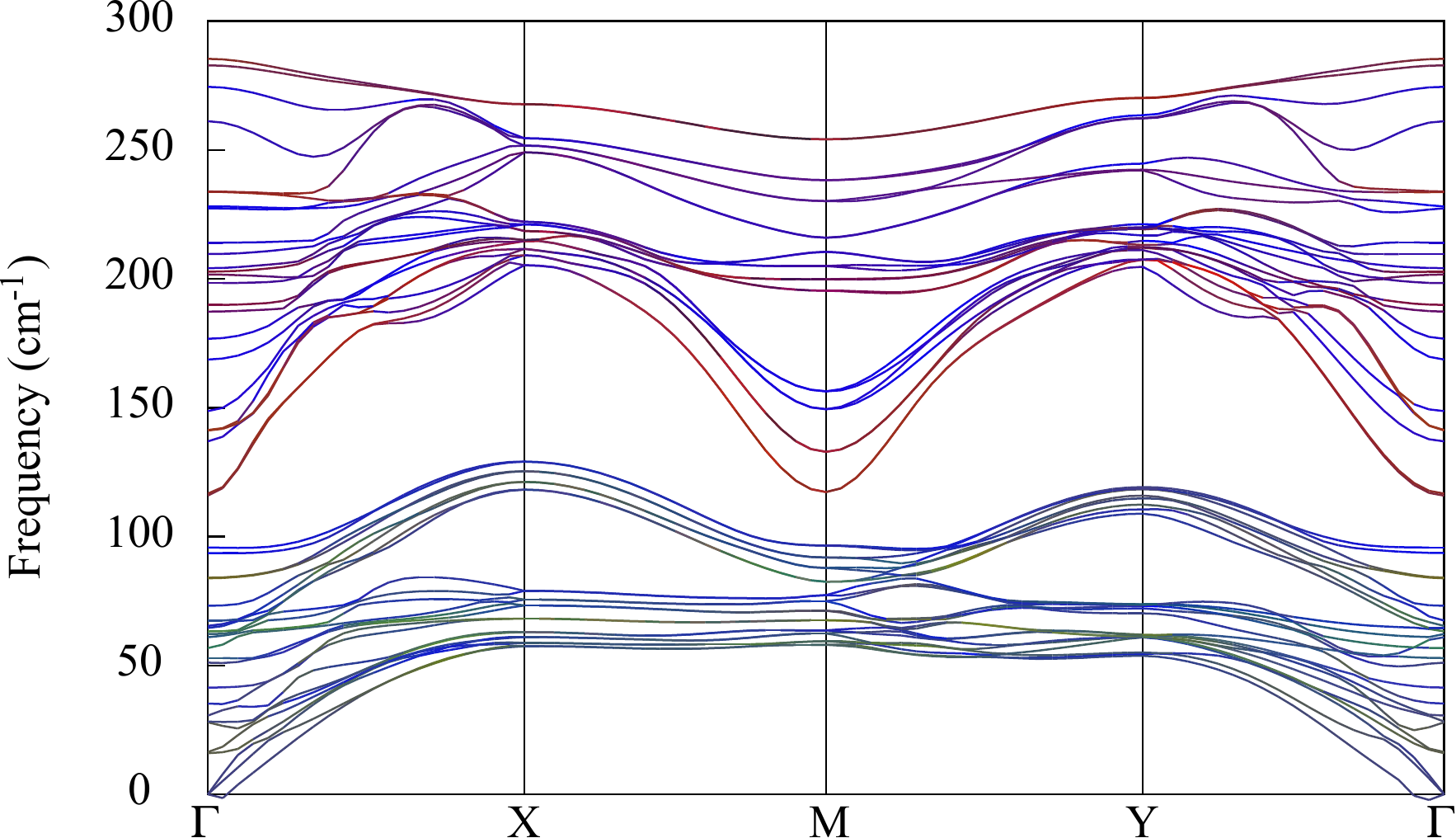}
\caption{Phonon band structure of a 4 layer SnS slab. Colors represent the contributions of the different atoms to the modes. Sn atom in green, S atoms from inner layers in blue, S atoms of surface layers in red. A mix of these three primary colors shows modes with contributions from different atoms. We identify a surface state at ~120 cm$^{-1}$ with lower energy than the corresponding bulk mode. Also, the two modes of highest energy involve only surface atoms.}
\label{phonons_SnS}
\end{figure}

Phonon dispersion curves were calculated for the bulk and the few layer compounds. In the monolayer case, we find a small unstable phonon mode at $\Gamma$ that breaks the 1/4-3/4 symmetry of the reduced coordinates in the X axis. In the other structures, the absence of any imaginary values confirms the dynamical stability of our predicted structures. The band structures split between low- and high-frequency manifolds. This is a signature of binary compounds where the masses of the two atoms are sufficiently different\cite{lindsay_2013}.
In SnS, the gap in the phonon dispersion curves is more pronounced than in SnSe, due to the larger difference in mass between the two atoms.

By projecting the mode eigenvectors over the atoms and identifying them with a color code (Fig. \ref{phonons_SnS}), we identify surface modes which are not present in the bulk compound. Indeed, the red curves in the phonon spectra represent modes from the S atoms at the surface. We clearly identify the red curves as modes where only S atoms from the surface dominate. At $\Gamma$, several surface modes are isolated from the bulk manifold, around 120, 140, 230 and 280 $cm^{-1}$.

\begin{figure}[htbp]
\centering
\includegraphics[width=0.5\textwidth]{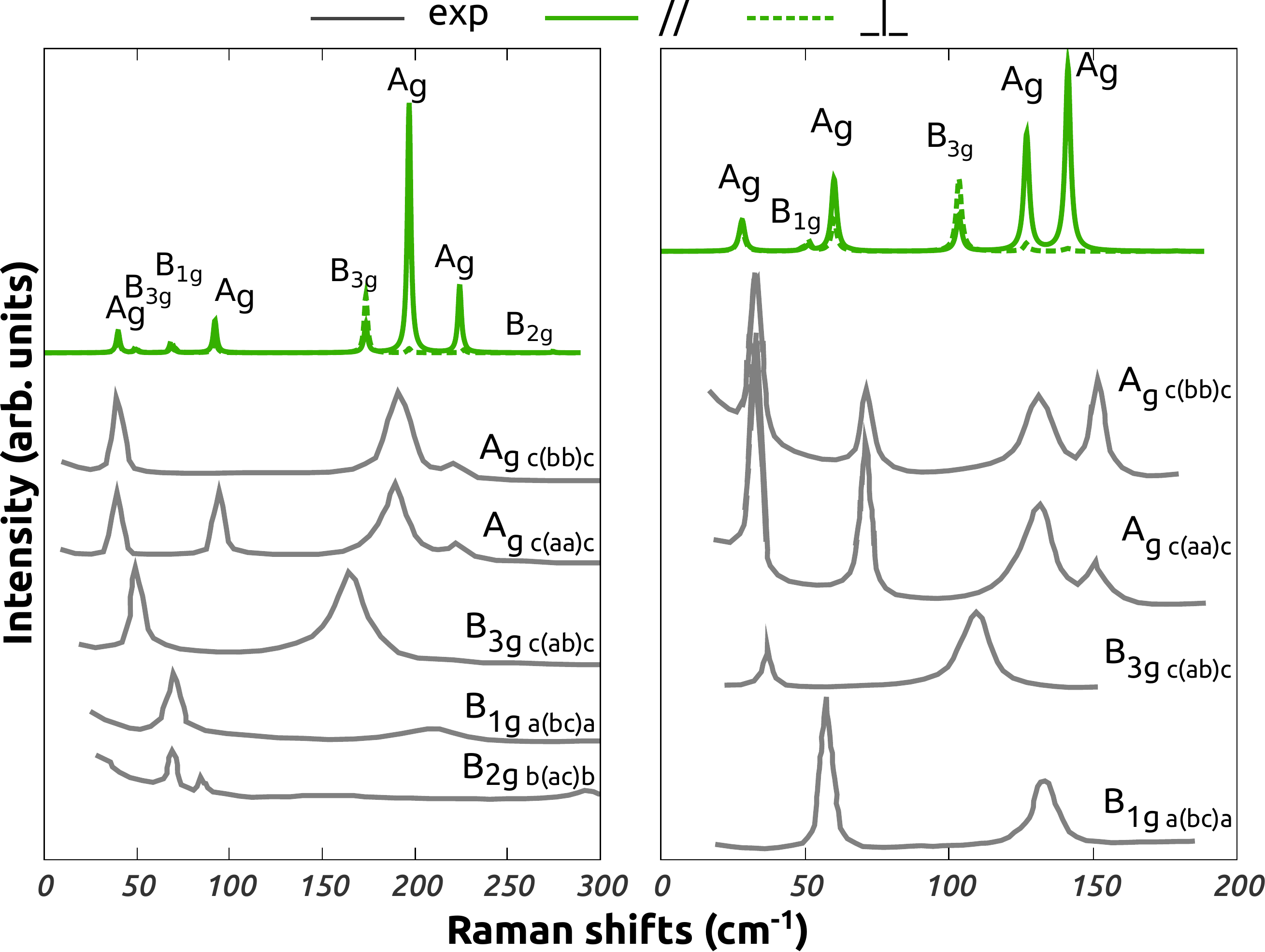}
\caption{Comparison of theoretical bulk results for SnS (left) and SnSe (right) with experimental work of Chandrasekhar \textit{et al.}  \cite{chandrasekhar_infrared_1977}. }
\label{raman_comp}
\end{figure}

\begin{figure}[htbp]
\centering
\includegraphics[width=0.5\textwidth]{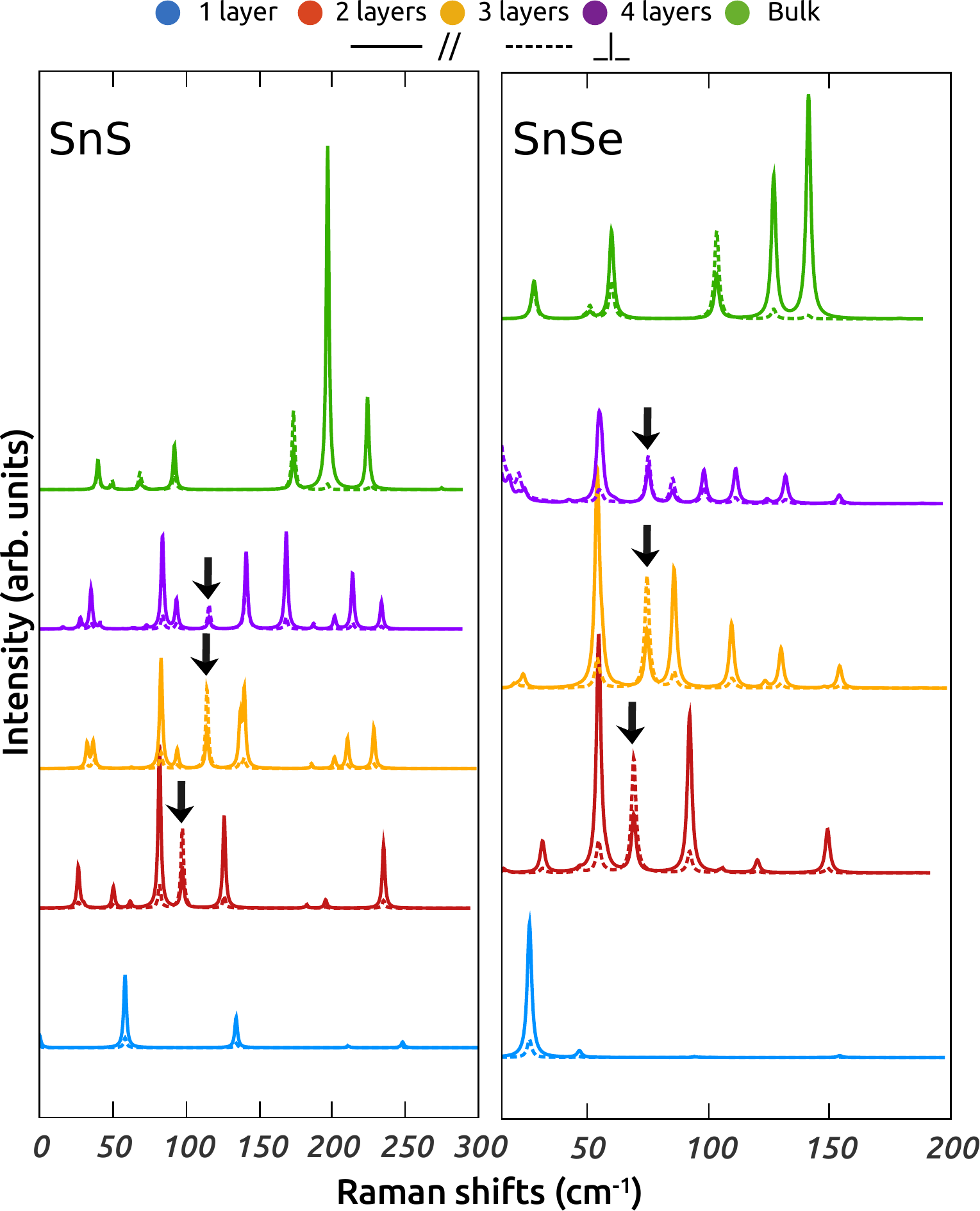}
\caption{Simulated powder averaged Raman spectra for SnX slabs. The surface modes, indicated by a black arrow, can be related to the B$_{3g}$ mode of SnS and SnSe bulk spectra at respectively 103 cm$^{-1}$ and 173 cm$^{-1}$ and shows a large parallel component compared to the other Raman active modes.}
\label{raman}
\end{figure}

A useful input for experimental characterisation is the prediction of Raman and reflectivity spectra. Fig. \ref{raman_comp} shows the comparison of our simulated Raman spectra for bulk compounds and the experimental results of Chandrasekhar~\textit{et al.} \cite{chandrasekhar_infrared_1977}, showing good agreement with calculated frequencies being within 13\% of the experimental ones. We also find agreement within 1\% between the thickest SnS flake presented in the work of  Li~\textit{et al.}  \cite{li_revealing_2017} and our theoretical study of the bulk compounds. B$_{1g}$ and B$_{2g}$ phonon modes do not appear in the experimental work of Li~\textit{et al.}  \cite{li_revealing_2017}. In our work, their relative Raman intensities is lower than the intensity of other modes. They might be hidden on a unified graph. 
Figure~\ref{raman} shows the evolution of the Raman spectra with the thickness. The Raman spectra change drastically between the 1- and 2-layer cases. There are fewer active peaks in the monolayer, due to its symmetry which is almost cubic. 
The surface states are also visible in the spectra as they are Raman active. These modes are indicated by a black arrow in Fig. \ref{raman}. The frequency of the surface modes appear to be converged already for 3- and 4-layer slabs. However, the relative intensity of the mode will most likely decrease with thickness, as can already be observed between 3 and 4 layers. We can relate the surface modes B$_{3g}$ of SnS and SnSe at respectively 103 cm$^{-1}$ and 173 cm$^{-1}$ by comparing their eigenvectors. The equivalence of the induced atomic displacements are shown in the supplemental information. This is also the only mode whose intensity in perpendicular polarization is much larger than for parallel polarization, giving an easy way to identify it, with a depolarization ratio $\rho = \frac{I_{\perp}}{I_{\parallel}}$ is larger than 0.75.

In supplemental information, we show the reflectivity of IR frequency waves normal to the surface, with electric field along the two optical axes of the crystal (as defined in Ref~\cite{gonze_1997}).
The different peaks correspond to polar phonon modes in resonance with the incident electric field.
In all spectra, the general shape is dictated by the resonance with one mode at about 225-250 cm$^{-1}$. More polar modes resonate when the number of layers increases. Also, the difference between the in-plane components of the IR reflectivity tensor evolves with the number of layers. For the 1-layer slab, the in-plane components are almost identical due to the square in-plane lattice parameters. However, with increasing number of layers, the difference between the X and Y component spectra becomes significantly noticeable as phonon modes of SnS (SnSe) with frequencies around 100 cm$^{-1}$ (75 cm$^{-1}$) resonate only with the electric field polarized Y axis. Tancogne-Dejean et al. demonstrated that standard ab-initio formalisms fail for the out-of-plane optical response in a 2D system \cite{tancogne_2015}, and we do not present here the Z component of the reflectivity.

\section{Born effective charges, Bader charges and Dielectric tensor}
We now turn to the dielectric response and the electronic density distribution of the SnX slabs.
The Born effective charge quantifies the variation of a material's polarization when the atoms are displaced and is defined as: 
\begin{equation}
Z^* = \frac{\partial^2 E}{\partial \mathcal{E} \partial \tau} = \frac{\partial P}{\partial \tau} = \frac{\partial F}{\partial \epsilon}
\end{equation}
with $E$ the total energy, $\mathcal{E}$ the electric field, $\tau$ an atomic displacement, $P$ the electrical polarization, and $F$ the force on the atom. The $Z^*$ also govern the frequency split between longitudinal and transverse optical modes.
We find that the Born effective charge tensors stay roughly constant as a function of the number of layers (see SI). This implies that the local electronic configuration of the atoms and the fundamental bonding nature do not change significantly when the number of layers varies.

To complete our analysis of the electronic density distribution, we also calculate the Bader charges \cite{bader_atoms_1994},
which confirm that the (static) electronic charge is not redistributed either by nanostructuring. The Bader charges increase slightly with the number of layers in the slab, showing a more ionic character in the bulk (see SI). 
As S is more electronegative than Se, the transfer of charge is larger in SnS compared to SnSe.

\begin{figure}[htbp]
\centering
\includegraphics[width=0.45\textwidth]{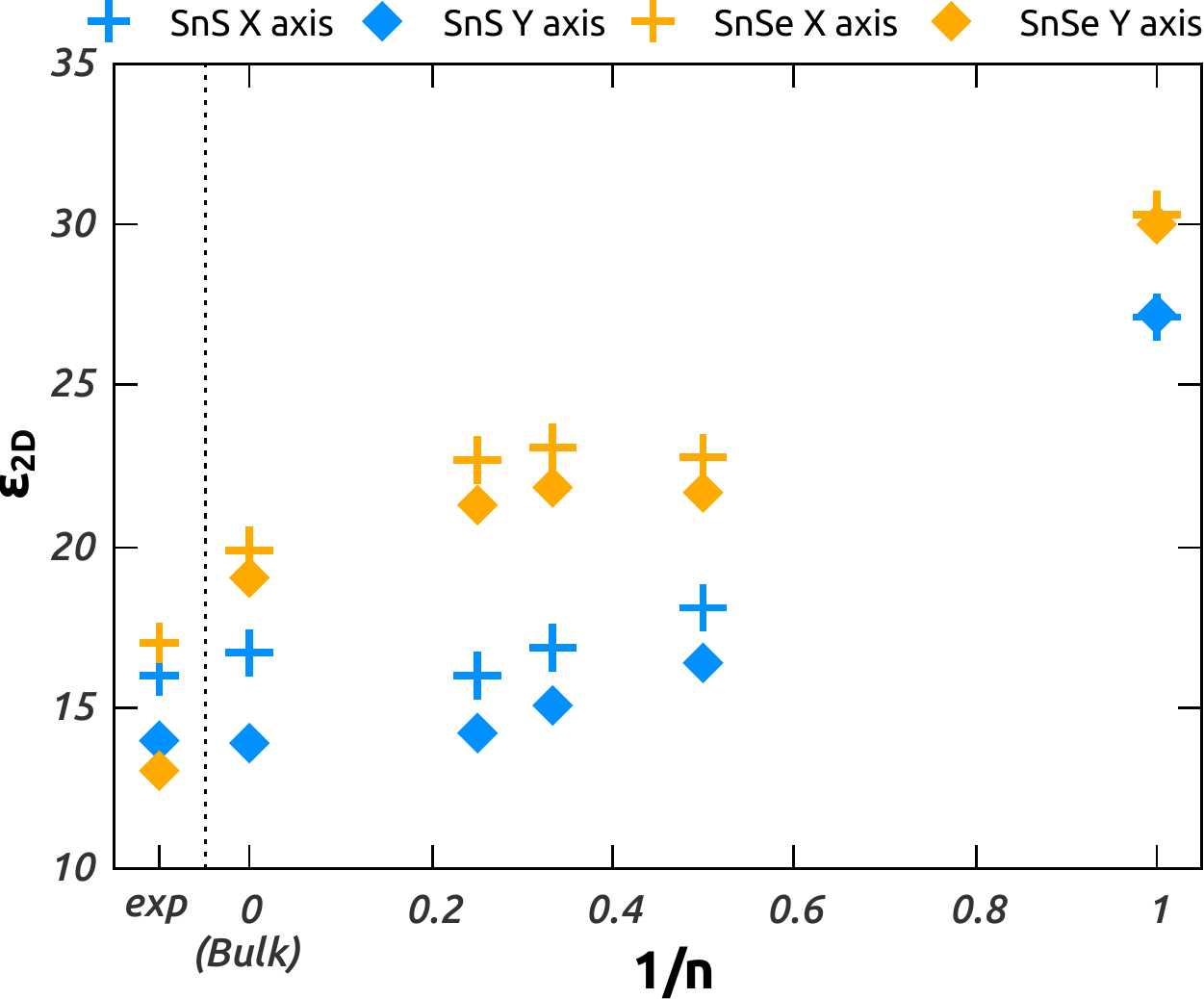}
\caption{In-plane components of the electronic dielectric tensor for SnS  and SnSe. The comparison with the experimental results of Chandrasekhar~\textit{et al.} \cite{chandrasekhar_infrared_1977} is shown on the left of the figure. }
\label{dielectric}
\end{figure}

\begin{figure}[htbp]
\centering
\includegraphics[width=0.45\textwidth]{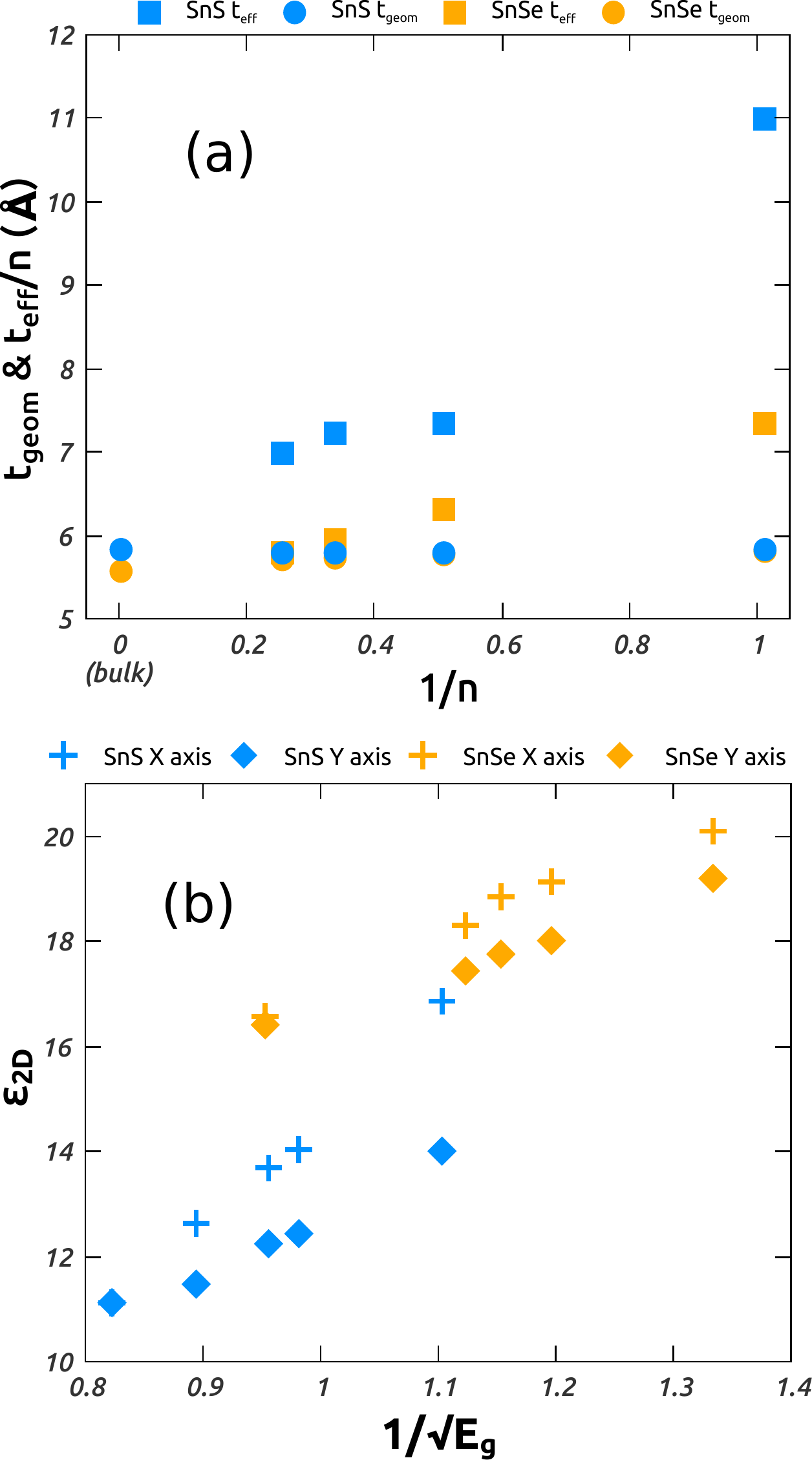}
\caption{(a) Comparison between $t_{eff}$ and $t_{geom}$. (b) Electronic dielectric constant $\epsilon_{2D}$ rescaled with $t_{eff}$ (note the subtle difference with Fig~\ref{dielectric}, as $t_{eff} \neq t_{geom} + \mathrm{constant}$)}
\label{dielectric_effective}
\end{figure}

Dielectric constant measurements are easy to perform and give crucial information on the electronic response of materials. Our bulk values (Fig. \ref{dielectric}) compare well with previous experimental works \cite{chandrasekhar_infrared_1977}. The computation of dielectric properties in 2D systems requires care as our calculations are performed with periodically repeated slabs separated by vacuum. In this periodic approach, the calculated dielectric tensor contains both the contribution of the slab and of the vacuum. For 2D materials of geometrical thickness $t$ computed in a cell of cross-plane lattice parameter $c$, an effective dielectric constant $\epsilon_{2D}$ can be derived as $\epsilon_{2D} = 1 + (\epsilon_{DFT}-1)\;\;{c}/{t}$ from the dielectric constant computed in the periodically repeated approach $\epsilon_{DFT}$\cite{freysoldt_screening_2008,gomes_phosphorene_2015,laturia_dielectric_2017,thygesen_calculating_2017}, to be able to compare the different 2D and 3D results. 
We choose the geometrical thickness $t_ {geom}$ as the distance between the two outermost atoms in a slab, plus one bulk interlayer distance - a common choice in Transition Metal Dichalcogenides (TMDs), where $\epsilon_{2D}$ is relatively insensitive to $t$. 
For both SnX compounds, however, the variation with $t$ is much stronger: 
the calculated effective electronic dielectric tensor $\epsilon_{2D}$, presented in Fig.~\ref{dielectric}, increases with 1/n. This behaviour is counter-intuitive, as $\epsilon$ should vary as $1 / \sqrt{E_g}$ \cite{Czaja1963}, and $E_g$ \emph{increases} with thinning. 

This effective dielectric constant model has been applied successfully on numerous systems including TMDs \cite{laturia_dielectric_2017} but has also been found to fail for monochalcogenides by Gomes and Carvalho in Ref.~\cite{gomes_phosphorene_2015}. In our case the gap decreases and the DFT bare dielectric constant increases with thickness, as expected, but the variations do not follow a square root law with a constant prefactor, and the model dielectric constant can be smaller in the bulk than in the monolayer.
In practice, this means that the effective dielectric thickness is not a simple function of the geometric thickness, and is super-linear for thin slabs. We have verified that this is not simply correlated to the extension of the electronic density outside the surface, which we find to be very similar in all slabs.
In Fig. \ref{dielectric_effective} (a), we present a comparison of the geometrical thickness $t_{geom}$ and an effective thickness $t_{eff}$ defined as the minimum thickness of the different layers required to retrieve the linearity between $\epsilon_{2D}$ and $1/\sqrt{E_g}$. In Fig.~\ref{dielectric_effective} (b), we also present the resulting electronic dielectric constant rescaled with $t_{eff}$, which recovers (by construction) the intuitive relation between $\epsilon$ and $E_g$.  
For the same reasons as for the reflectivity, we do not report the out-of-plane component of $\epsilon$.

\section{Summary \& Conclusions}
We calculate electronic and phononic properties of SnS/SnSe slabs from 1 to 4 monolayers in order to pinpoint spectroscopic signatures of 2D material thickness. We propose several simple non-invasive techniques to discriminate mono and few layer Sn chalcogenide samples. The behaviour of these two iso-electronic compounds is globally similar upon nanostructuring, but presents subtle differences. We identify structural distortions of the unit cell, as the thickness of the slab is reduced, leading to an almost cubic symmetry for the monolayers. The structural evolution of the unit cell with nanostructuring drives the transition from indirect to direct band gap as the number of layers is reduced, and the optical band gap expands. 
Surface phonon modes are identified by projecting the phonon bands over the atoms. They can be associated to a B$_{3g}$ bulk mode through a comparison of the eigenvectors of the respective phonon modes. The unique depolarization ratio of these modes allows us to identify them in the Raman spectra. Furthermore, the monolayer spectra shows a specific feature : the number of active modes are reduced compared to thicker layers, which can then be used to distinguish between a monolayer, a few-layer slab or a thicker sample. Reflectivity spectra show a similar evolution and can also be used to determine the thickness.
The local electronic environment of the atoms are almost independent of layer number, as quantified by the Bader (static) and Born (dynamical) charges. 
Finally, the electronic gap and phononic properties have a strong thickness dependence, and common models for the effective 2D dielectric constant break down, due to a super-linear variation of the effective dielectric thickness.
The results have the potential to enable fast recognition of ultrathin chalcogenide samples, and we hope to stimulate experimental work on the dielectric properties of these systems.

\section{Methods}

Density Functional Theory (DFT) calculations are performed using the ABINIT package \cite{Gon-05, Gon-09}, which implements the plane-wave methodology (here using norm-conserving pseudopotentials). The exchange-correlation energy is given by the generalized gradient approximation (GGA) of Perdew, Burke and Ernzerhof \cite{perdew96}. Norm-conserving Troullier-Martins type pseudo-potentials generated with fhi98PP code are used to describe interactions between atomic cores and valence electrons of SnSe and we use an ONCVPSP \citep{hamann_2013_oncvpsp} generated pseudo-potential for SnS which produces lattice parameters that compare well with experiment \cite{Zhao_2014,chandrasekhar_infrared_1977}.

We have checked that including Van der Waals interactions within the Grimme D3 approximation \cite{grimme_consistent_2010} does not affect the interlayer distance significantly, and it is not employed for the results above.

The wave functions  are represented in a plane-wave basis with a cutoff energy of 30 Ha for SnSe and 40 Ha for SnS. The reciprocal space of SnSe bulk is sampled with a $4 \times 4 \times 4$ Monkhorst-Pack-type grid  \cite{Monkhorst_Pack_1976}, whereas  an $8 \times 8 \times 8$ unshifted grid is used for SnS bulk. 
The total energy is converged to within 3 meV per unit cell with respect to the k-point grid and cutoff kinetic energy of the plane waves used as a basis set. Atomic positions and lattice parameters are relaxed using a Broyden \cite{broyden_1965} algorithm until the maximal absolute force on the atoms is less than $10^{-6}$ Ha/Bohr. The phonon band structure along high symmetry lines is obtained by standard methods based on response function calculations and DFPT \cite{verstraete_2014_dfpt}. Ten irreducible q-points from an unshifted $4 \times 4 \times 4$ grid are used for the calculation of dynamical matrices. Electron band structures are calculated on a finer $24 \times 24 \times 24$ k-point grid. For few layer calculations, convergence with respect to the size of the vacuum in the unit cell is also performed, to within 1 meV with a vacuum gap of 20 Bohr. Also, the dispersion along the vacuum is suppressed by considering k-point and q-point grids with only one point along the Z axis.

Raman intensities are computed using perturbation theory to calculate the third derivative of the energy with respect to two electric fields and one atomic displacement \cite{Veithen_2005}. In our simulations, the energy of the incident light is chosen to be 2.41 eV (514,8 nm) and the temperature 300 K~ \cite{Chandrasekhar_1977}. We use a common approximation of calculating the third derivative only within the local density approximation. The width of the Raman peaks is mainly determined by anharmonic scattering, which limits the phonon lifetime. We do not consider this effect here, and broaden the peaks with a Lorentzian function having a fixed width of $5 ~ 10^{-6}$ Ha. The Raman tensor is averaged to represent a powder sample as in Ref.~ \citep{Caracas_2010}.

\acknowledgments
The authors gratefully acknowledge funding from the Belgian Fonds National de la Recherche Scientifique FNRS (PDR T.1077.15-1/7) and the Communaut\'{e} Fran\c{c}aise de Belgique (ARC AIMED 15/19-09). Computational resources have been provided by the Consortium des Equipements de Calcul Intensif (CECI), funded by FRS-FNRS G.A. 2.5020.11; the Tier-1 supercomputer of the F\'ed\'eration Wallonie-Bruxelles, funded by the Walloon Region under G.A. 1117545; and by PRACE-IP DECI grants, on ARCHER , Beskow and Salomon (ThermoSpin, ACEID, OPTOGEN, and INTERPHON 3IP G.A. FP7 RI-312763 and 13 G.A. 653838 of H2020). ZZ acknowledges financial support by the Ramon y Cajal program (RYC-2016-19344), the Spanish program PGC2018-096955-B-C43 (MICIU/AEI/DEDER,UE) the CERCA programme of the Generalitat de Catalunya (grant 2017SGR1506), and by the Severo Ochoa programme(MINECO, SEV-2017-0706). We are also grateful to the Royal Society, the European Research Council (ERC-2009-StG-240500-DEDIGROWTH), and the Engineering and Physical Sciences Research Council block grants for their financial support.

\bibliography{SnS}

\end{document}